\def\year{2021}\relax
\documentclass[letterpaper]{article} 
\usepackage{aaai21}  
\usepackage{times}  
\usepackage{helvet} 
\usepackage{courier}  
\usepackage[hyphens]{url}  
\usepackage{graphicx} 
\usepackage{natbib}  
\usepackage{caption} 

\usepackage{amsmath}
\usepackage{amssymb}
\usepackage{braket}

\title{Computability, Complexity, Consistency and Controllability: A Four C's Framework for cross-disciplinary Ethical Algorithm Research}
\author{
    Elija Perrier\textsuperscript{\rm 1}\textsuperscript{\rm 2}\\
}
\affiliations{
    \textsuperscript{\rm 1}Centre for Quantum Software and Information\\

    University of Technology, Sydney\\
    Sydney NSW 2000\\
    
\textsuperscript{\rm 2} Humanising Machine Intelligence Program \\The Australian National University, Canberra\\
elija.t.perrier@student.uts.edu.au
}

\begin{document}

\maketitle

\begin{abstract}
The ethical consequences, constraints upon and regulation of algorithms arguably represent the defining challenges of our age, asking us to reckon with the rise of computational technologies whose potential to radically transforming social and individual orders and identity in unforeseen ways is already being realised. Fittingly, concurrent with the emergence of such epoch-shaping technologies has emerged a rapidly expanding and multi-disciplinary set of research disciplines focused on these very questions. As the inexorable march of computational technologies encroaches across society and academic disciplines, it is natural that diverse specialisations including computer science, moral philosophy, engineering, jurisprudence and economics should turn their attention to the algorithmic zeitgeist. Yet despite the \textit{multi}-disciplinary impact of this \textit{algorithmic turn}, there remains some way to go in motivating \textit{cross}-disciplinary collaboration is crucial to advancing feasible proposals for the ethical design, implementation and regulation of algorithmic and automated systems. In this work, we provide a framework to assist cross-disciplinary collaboration by presenting  a `Four  C’s Framework' covering key computational considerations researchers across such diverse fields should consider when approaching these questions: (i) computability, (ii) complexity, (iii) consistency and (iv) controllability. In addition, we provide examples of how insights from ethics, philosophy and population ethics are relevant to and translatable within sciences concerned with the study and design of algorithms. Our aim is to set-out a framework which we believe is useful for fostering cross-disciplinary understanding of pertinent issues in ethical algorithmic literature which is relevant considering the feasibility of ethical algorithmic governance, especially the impact of computational constraints upon algorithmic governance. 
\end{abstract}

\noindent The rapid acceleration of artificial intelligence and algorithmically-based technologies over the last several decades has led to an explosion in research programmes dedicated to framing, understanding and regulating the ethical, moral, philosophical and jurisprudential consequences of these powerful and epoch-defining technologies. As computational and information sciences advance throughout all facets of social, economic and political dimensions of society, it is natural that responses to this technological imperative involve multidisciplinary research fields, including the burgeoning fields of fair machine learning \cite{chouldechova_fair_2017, caton_fairness_2020}, algorithmic jurisprudence \cite{zalnieriute_rule_2019}, interpretability and regulatory technology. The rise of such technologies and, moreover, the increasing envelopment of technical and epistemological practice by algorithmic technologies, a phenomenon we denote as the \textit{algorithmic turn} (in homage to its linguistic namesake), is as profound as it is ubiquitous. The extraordinary reach of such technologies naturally motivates cross-disciplinarity in such research programmes among ethicists, philosophers and computer scientists in order to grapple with the consequences of these technologies. While such cross-collaboration has increased over several years, there remain challenges in translating between the epistemological citadels within which each discipline is resides. Though multidisciplinary in impact, research programmes often remain less cross-disciplinary than called for. An example is within fair machine learning research, where technical results on the theory and mathematics of implementing fairness criteria algorithmically (to ameliorate bias, discrimination for example) tend to have limited, if any, engagement with the directly relevant and vast literature within ethics, social choice and decision theory, philosophy and jurisprudence. 

Of course, disciplinary silos are nothing new in the academe and specialised focus within disciplines is critical to pushing frontiers of research. Moreover, it is impossible for there to be a single, overarching, epistemological framework to provide direction to such a variegated topic as ethical algorithms. Nevertheless, as we contend in this work, there are benefits to be had by providing a framework where researchers in the sciences, economics and humanities can engage with common themes, protocols and domain criteria relevant to ethical consideration of algorithms. 

For example, seeking to ethically constrain or shape algorithms via design or regulation requires domain understanding of precisely what is meant by the term \textit{algorithm}, along with familiarity with core concepts in computational science which inform algorithms themselves, such as computational complexity, computability and others. Similarly, programmes that seek to explore ethical constraints upon algorithms would benefit considerably from a consideration of how issues and dilemmas facing their field, such as the inevitability of trade-offs among fairness criteria, probabilistic approaches to ethical classification and auditing have been explored in fields such as moral philosophy, population ethics and decision theory. Indeed the very validity of fairness criteria themselves, for example, is a rich and ancient debate across ethics and associated disciplines. 

Our answer to this challenge is to propose a framework based upon core computational science concepts of computability, complexity, consistency and controllability, which we name the `Four C's Framework'. The aim of the framework is to set-out core concepts from computer science applicable to consideration of the ethical status and consequences of algorithms and automated systems, in order to provide researchers across disciplines with a series of considerations that we believe should be taken into account in any proposals for ethical regulation and assessment of algorithmic systems. Though care has been taken to address the nuances of various research fields, inevitably any cross-disciplinary synthesis that seeks to abstract key concepts will simplify, gloss over and omit any number of important results.

\subsubsection{Results and Contributions}
The contributions of our paper relative to the state of the art are as follows:
\begin{enumerate}
    \item Development of a holistic framework to assist cross-disciplinary collaboration among researchers by setting out key technical research criteria concerning computability, complexity, consistency and controllability which should be taken into consideration when the ethical classification, regulation or assessment of algorithmic systems is undertaken.
    \item Explication of connections among computational science, philosophical and economic science research in a way that identifies common problems both seek to address and that assists in motivating the translation of results in one field, say social choice and decision theory, to another, such as fair machine learning.
\end{enumerate}

\subsubsection{Structure}
The structure of our paper is as follows. \textbf{Part I} sets out a framework for understanding how to classify and characterise algorithms as ethical. It introduces conceptualisations including ethical computation, ethical decision (deontological) procedures as distinct from ethical (consequentialist) outcomes. We connect ethical auditing to provability concepts and argue algorithmic governance in the limit will be itself necessarily algorithmic. \textbf{Part II} of the paper introduces the Four C's Framework described above. It unpacks concepts from computability and complexity theory, such as efficiency and feasibility as they pertain to ethical algorithmic analysis. The importancy of consistency (and completeness which we roll into consistency considerations), including the extent to which consistency in process or outcomes are ethically significant, is examined. We similarly explore how controllability is a critical element to consider when assessing ethical algorithms. We conclude the part with consideration of the ethical consequences for the inevitable reliance upon heuristics in any algorithmic regulatory context. \textbf{Part III} focuses on cross-disciplinary connections between algorithmic epistemology and other disciplines, such as impossibility results in social choice and social welfare theorems, contextualising population ethics' debates among other examples. \textbf{Part IV} concludes our paper, exploring a number of cross-disciplinary options for handling key challenges with ethical algorithms, including determining appropriate trade-offs, uncertainty in algorithmic processes or outcomes, via analogy with existing institutions, such as the law, which face similar challenges.

\section{Part I: Ethical Algorithms}
\subsection{Ethical algorithms}
In this section, we propose a framework for understanding the intersection of ethics and computational science in order to provide a basis from which to assess how quantum computation and information processing impacts research into the ethics of algorithms and automated, computational, systems. Our guiding presumption is that ethical governance of automated systems, artificial intelligence and algorithms fundamentally requires an understanding of (i) the nature of computation and, importantly, (ii) an awareness of the limitations of what is possible computationally. Research programs on ethical algorithms must reckon with a range of important considerations, including: computational constraints on how algorithms may be regulated, the impact of impossibility theorems (such as in social choice theory \cite{arrow_difficulty_1950}), population ethics' dilemmas \cite{parfit_reasons_1986}. Important concepts from logic and computer science, including the appropriate formal logics and languages within which ethical judgments are made, the relationship of canonical concepts such as provability, consistency and completeness to the theory and application of ethical algorithms are important. So too, the ways in which computational probability and uncertainty intersects with similar concepts in formal ethical theories, such as the vast literature on decision theory \cite{savage_foundations_1972} are crucial. How consistent must ethical theories or conclusions be? What thresholds, consistencies or resource-dependencies (such as bounded rationalities) are acceptable for different algorithmic systems? Beyond the philosophical and technical, ethical algorithmic science must consider legal and jurisprudential contexts governing the use and role of such technology in social contexts.\\
\\
In this work, we proceed under the conjecture that the complexity (both formal and in terms of size and feasibility of auditing) of both \textit{data} and computational \textit{systems} will necessitate ultimately the \textit{regulation of algorithms by algorithms}. We argue that this entails that ethical governance must be framed within and is constrained by computability and complexity constraints, which naturally leads to consideration of the role of \textit{heuristics} (in a formal computational sense) and the ubiquity and normativity of probabilistic reasoning. In particular, we argue that this implies - as has been explored throughout the diverse literature - the potential and inevitability of inconsistent ethical outcomes whose reconciliation (namely, which ethical criteria to prefer) itself is dependent upon decision-procedures themselves amenable to algorithmic encapsulation. It also engenders its own set of ethical dilemmas and decision-problems regarding uncertainty over applicable or competition among applicable ethical theories and, therefore, whether an algorithm or algorithmic outcome is properly classified as ethical. This in turn leads to the well-explored consideration of the necessity of unavoidable ethical trade-offs \cite{dwork_fairness_2012,kearns_empirical_2019}. Moral theories governing the ethics of computation must reckon, then, with the \textit{type} of inconsistencies that are ethically permissible. We argue that in turn, such computational constraints impose on regulatory architects and policymakers a need to consider computational limitations when framing and implementing ethical regulation of algorithms and artificial intelligence.

\subsubsection{What is an ethical algorithm?}
What constitutes an ethical algorithm, ethical artificial intelligence or ethically satisfactory computational or automated systems has a rich and diverse provenance across a range of fields including automated engineering disciplines, decision theories and formal moral reasoning. Firstly, and importantly, we take as our basic model of computation the canonical Turing machine i.e. one that can simulate an arbitrary (classical, we omit reference to quantum algorithms, a discussion of which is set-out in separate work in progress) computations. Thus we define as an algorithm in the usual way of computational science, as any sequence of operations simulable using a Turing-complete system \cite{sipser_introduction_2012}, including typical recursive functions. We assume that no matter the computational device or intricacy, such a device ultimately realises the Turing model of computation - and that references to algorithmic or automated computation refer back accordingly. 

Below we set out a framework which we believe is useful for not only consideration of ethical algorithmic questions, but one that connects to consideration of the ethics of quantum computation.
\begin{enumerate}
    \item \textit{Ethical computation}. We define ethical algorithmic computation as consisting of: (i) \textit{ethical decision procedures} by which we mean decision-procedures where the process of computation itself satisfies some ethical criteria or set or rules which is provably ethical according to a formal proof system. In this sense, ethical decision procedures consider \textit{ethical algorithmic means} or, more formally, are a species of \textit{algorithmic deontology} concerned with normative theories regarding which algorithmic processes are morally required, forbidden, or permitted according to a meta-algorithmic rule; (ii) \textit{ethical outcomes} or what we denote as \textit{algorithmic consequentialism}, where the consideration is whether the outcome of an algorithm or computation is provably ethical. In this sense, the consideration here intersects with extensive debates over consequentialism \cite{pettit_consequentialism_1993, sinnott-armstrong_consequentialism_2019}. Both definitions here imply a conception of both computation and algorithms, for which we appeal to the standard computational conceptions of Turing machines and algorithmic effective procedures \cite{sipser_introduction_2012, mendelson_introduction_2015}.
    \item \textit{Auditing and provability.} The next consideration is that for a computation or algorithm to be ethical entails in the strict sense that its status as an ethical computation is \textit{provably} ethical according to the relevant criteria above. This in turn intersects with concepts of provability (in formal systems see \cite{boolos_logic_1995}). An example in ethical algorithmic contexts is research into auditing of measures of statistical fairness \cite{kearns_preventing_2018}. What counts as \textit{provably} ethical is both philosophically contested (the appropriate criteria, formal system to adjudge ethical status by) and subject to computational constraints, including whether a solution is obtainable (i.e. is it possible to find an algorithm satisfying deontological or consequentialist criteria given the computational resources available e.g. is a solution in a too-hard complexity class?) and, if given a solution, is it recognisable \textit{as} an ethical algorithm such that one may \textit{prove} that given an algorithm, it meets such criteria. Auditing is also an important consideration for \textit{interactive} algorithms, by which we mean algorithms whose computational structure or output is contingent on responses from the environment. Often it is assumed that the act of ethically vetting an algorithmic system is independent of the implementation of that algorithm itself. Yet even existing examples demonstrate that algorithms, particularly those with the ability to `self-program' or dynamically adjust their path (such as via policy updates in reinforcement learning contexts \cite{sutton_reinforcement_2018}) opens up ethical algorithmic analysis to entirely new fields that consider both strategic behaviour (such as the game theoretic behaviour of ethical algorithms under auditing) and provability and information theory (such as how one can reliably prove an algorithm has satisfied ethical criteria, or whether an algorithm is masking its unethical activities or when to trust auditing algorithms themselves). In turn, such considerations connect with the diverse and important fields of information theory, cryptography and other areas in information science. We note that the simplistic process/outcome (or means/ends) distinction itself can be recast in terms of an assessment of Turing formalism, nevertheless we retain it for pedagogical purposes.
\end{enumerate}
 Ethical algorithmic problems are \textit{complex} and challenging thus both from a computational perspective but also epistemological perspective, including the appropriate ethical theories to apply. Pragmatically, high volume, high velocity and highly diverse datasets present an immediate challenge in terms of scalability (the curse of dimensionality) while the complexity and opacity of algorithmic processes problematise their interpretability \cite{lo_piano_ethical_2020, doshi-velez_towards_2017}. This is already apparent across a wide variety of use-cases in fields such as finance, communications, cybersecurity, autonomous machines and complex and dynamic code structures (e.g. how to monitor millions of neural networks at any one time). As we discuss in more detail below, the complexity and uncertainty of both data and computational processes is a crucial feature to be considered in any approach to ethical algorithmic regulation. In particular, we conjecture that the complex nature of algorithmic systems, large datasets and the ubiquity of artificial intelligence (or even simply automated computational systems) will necessitate that, by and large, ethical governance (especially auditing and procedural regulation to check satisfaction of ethical criteria) will itself need to be via algorithmic computational means. The use of algorithmic auditing is itself commonplace across diverse engineering fields for example, which deploy automated systems in order to sample, test and verify the veracity of such systems (for example auditing circuit manufacture or aircraft engineering). Thus we argue that research programmes focused on regulation of algorithms must contend with the reality that the complexity of the computational and algorithmic landscape necessitates \textit{algorithms regulating algorithms}.
 
 \section{Part II: Four C's Framework}
 Here we present the \textit{Four C's Framework}: (i) computability, (ii) complexity, (iii) consistency and (iv) controllability, a framework which we believe is useful for considering the feasibility of ethical algorithmic governance, especially the impact of computational constraints upon algorithmic governance. This framework has application to the governance of both classical and computational systems and data. We explain each of these concepts in turn.
 \begin{enumerate}
     \item \textit{Computability.} Two questions that must be asked when considering the ethics of computation, especially the ethical classification (deontological or consequential) of algorithms are: (i) is the ethical computation (process or outcome) actually \textit{computable}?; and (ii) if so, is the ethical computation \textit{efficiently} or \textit{feasibly} computable. Here we distinguish between \textit{efficient} ethical computations, where efficient refers to the standard computational definition \cite{papadimitriou_computational_1994, boolos_computability_2007} but also \textit{feasibility} that is, is the computation (including the computation necessary to classify as ethical, or audit an algorithm as such) able to be undertaken given resource constraints? And indeed, is the computation feasible within the time-scales or constraints given by context: an ethical computation may well be tractable (computable in polynomial time), yet take such a long time to compute that by human time-scales or time-scales required for decisions to be made, it is \textit{effectively intractable}. For algorithmic governance, computability considerations mean considering how regulation must provide a decision-criteria for the determination of the status of an algorithm as deontologically or consequentially ethical. \\
     \\
     The types of considerations will include: (a) \textit{decidability}, whether ethical criteria are well-defined, consistent and decidable? One must know whether an algorithmic procedure or output can be definitively classified as ethical or not. One must ask also whether there is a decision procedure to decide among competing ethical algorithms (for example, is there a decision procedure for deciding between which ethical settings a driverless car should adopt); (b) \textit{deterministic v. probabilistic}, is the assessment of ethical status deterministic or probabilistic? Can algorithmic governance be ethical if its ethical status is not computable, uncertain or probabilistic? This is of particular relevance to the quantum computational context in which the specification of both quantum information processes used in an algorithm may be unknown \textit{and} because the outcome of quantum computations is ultimately a probabilistic distribution \cite{nielsen_quantum_2010}; and thirdly (c) \textit{uncertainty/risk thresholds}, that is, if ethical classification of algorithms or their output is probabilistic (as it necessarily is in most conceivable quantum algorithmic cases), how are decisions around acceptable risks of unethical outcomes determined? Do these decision procedures themselves need to satisfy some sort of logical or ethical constraints?
     \item \textit{Complexity.} Computational complexity is a cornerstone of computer science which considers how the resources needed to computationally solve a problem scale with the input size $n$ to the problem \cite{papadimitriou_computational_1994,aaronson_why_2011}. An algorithm (ethical or not) is considered (i) \textit{efficient} if its run-time is upper-bounded by a polynomial function of $n$, in which case it is solvable by classical computation; and (ii) \textit{inefficient} if the run-time is lower-bounded by an \textit{exponential} function of $n$, rendering the problem in a higher complexity class (e.g. EXPTIME) such that it cannot be efficiently solved by classical computation. As mentioned above, within this category we also include infeasibility, where an algorithm is infeasible if it is efficient, but its resource needs exceed the available computational resources required to carry it out e.g. when a computation takes too long relative to some desiderata. Complexity considerations are crucial to ethical algorithmic science, such as in fair machine learning contexts on complexity of decision-trees  \cite{valdivia_how_2021, fitzsimons_intersectionality_2018} and auditing subgroup fairness \cite{kearns_preventing_2018} as two examples. Complexity is an important consideration to ethical algorithms because it is an important consideration for algorithmic theory overall: most problems are not tractable, e.g. the size of higher-order complexity classes vastly outstrips that of $P$ and $NP$ for example.
     \item \textit{Consistency.} Computational and ethical algorithmic consistency is the third category of our framework and concerns the extent to which ethical algorithmic decisions and procedures are \textit{consistent}. By consistency, we draw upon standard concepts of logical consistency (i.e. that a theorem is not provably true and false within a logic, and that a computation will not output a statement as true and false). More particularly, we expand consistency in an algorithmic context as follows to: (i) \textit{process (deontological) consistency}, namely are the decision-procedures of the algorithm consistent? Do similar decisions follow similar methodologies? Such considerations are important, for example, in various fields of fair machine learning \cite{caton_fairness_2020} and also concepts of representational justice (such as the rights of an individual to have or not have their representations construed in particular forms algorithmically); and (ii) \textit{outcome (consequentialist) consistency}, namely whether outcomes are consistent. For example, must two algorithms making ethical decisions, say classification into the same class, come to the same conclusion (especially of probability or stochasticity in general is a factor)? Must the set of all ethical algorithmic decisions be consistent with all ethical norms at all times? If not, how does one decide what to include or omit from ethical criteria? Research questions on consistency should also consider questions about \textit{maximal consistency}, that is, whether ethical classification must form a maximal consistent set, in some sense must classification as ethical include the maximum number of classifications possible which are also consistent? In this sense, we are asking what in logical and computational contexts is the extent to which an ethical classification regime needs to be \textit{complete} such that it can properly classify all algorithms.
     
     Another important consideration around the topic of consistency for ethical algorithmic governance is what we describe as the proscribed limits of algorithmic computation due to consistency consequences. Put colloquially, does society wish to normatively constrain the extent to which algorithms may reveal \textit{inconvenient truths} which expose the inherent inconsistency or contradictions in accepted normative criteria, that is, reveal potential impossibility results? As a concrete example, fair machine learning contexts commonly proceed upon the assumption of maximising utility for or minimising disutility for a protected class $S$ of persons. Such assumptions are themselves based on particular ethical, moral and normative (logical) soundness of the utility and validity of designating $S$ as a protected class. Yet it is entirely possible that an algorithmic approach to ethical determinations may reveal such assumptions to be unsound (by, for example, revealing the reasoning according to which they were deemed sound to be invalid) or inconsistent with other ethical theorems, such as normative assumptions about social welfare utility. Ethical algorithmic research must therefore contend with the limits and frontiers of how far algorithms may probe the assumptions and normative axioms for a given context.

     \item \textit{Controllability.} The final category is that of controllability, which considers whether and under what circumstances an ethical algorithm is \textit{controllable}. In this context, ethical algorithmic research can draw upon the rich literature of classical (and quantum) control across engineering, mathematical and physics disciplines. Questions that should be posed for ethical algorithmic consideration include: (a) \textit{reachable controls}, are controls available to steer an algorithm to a reachable ethical state (e.g. steering it towards an ethical outcome or a guaranteed use of ethical processes in computation)? Researchers can draw upon concepts from classical control theory which considers whether a particular state is reachable given the set of controls available \cite{dorf_modern_2011}; (b) \textit{ethical controls}, questions explore the types of controls which are ethically appropriate. To what extent must, from the perspective of ethical governance, control mechanisms themselves satisfy ethical criteria? For example, (i) `black box' methods vary inputs with respect to some desired output state, while the details of the computation are unknown (akin to certain neural network architectures or more specifically quantum computational processes); and (ii) `open box' methods where control over (arbitrary) aspects of the computation are permissible; and (c) \textit{open (offline) v. closed loop (online) control}, which considers the extent to which humans should be `in the loop' for controlling quantum systems \cite{gerdes_implementable_2016} (with trade-offs in efficiency and accuracy), versus `closed-loop' (online) control, where questions about trust, accuracy and moral agency are raised; and finally (d) \textit{noise}, how does the presence of noise (errors or uncertainty in data or outputs) affect the ethical control of algorithms (e.g. fairness under measurement error \cite{liu_delayed_2018}).
 \end{enumerate}
 
 \subsubsection{Necessity of heuristics}
 The challenges to ethical governance posed by computational and ethical considerations illuminated by a Four C's analysis as outlined above, such as complexity constraints, are not themselves necessarily new or constrained to computational science. Human society as a complex system has evolved, principally via its legal, technological and ethical social systems, to organise, order and function in the face of high degrees of system complexity. As a result, ethical algorithmic governance regimes need to countenance the following consequences.
 \begin{enumerate}
     \item \textit{Computational complexity necessitates heuristics.} The complexity (tractability and feasibility) limits upon ethical algorithms mean that heuristics are a necessary part of ethical governance: (a) \textit{ethical computation inefficient}, it may be that there is no computationally efficient algorithm for undertaking a particular ethical algorithm or algorithm ethically (e.g. see \cite{kearns_preventing_2018} on subgroup fairness); (b) \textit{computationally infeasible}, it may be that regardless of computational efficiency, complexity may grow rapidly so as to render ethical computation or its generalisation to novel contexts infeasible (e.g. where there is a high-degree polynomial runtime) given available resources; (c) \textit{probabilistic ethics}, the use of heuristics or non-deterministic solutions can render ethical computational claims uncertain or probabilistic, which in turn may necessitate the use of heuristics thresholds for example.
     \item \textit{Heuristic solutions necessitate ethical trade offs.} As is well-established across a variety of branches of ethical computation (see \cite{dwork_fairness_2012}), constraints upon ethical computation can expose the inherent inconsistency of ethical imperatives and require risk-assessment of ethical trade-offs. Consequences for this include the following: (a) \textit{probability approximately ethical}, is algorithmic learning of ethical labels approximate and if so what error rates are acceptable? An example is probably approximate metric-fair learning \cite{yona_probably_2018}; (b) \textit{choice of bias}, the choice among algorithms is unlikely to be between unbiased and biased but rather an acceptable choice of bias (see \cite{kleinberg_inherent_2016}), yet this itself requires a further decision-procedure for determining acceptability; and (c) \textit{heuristic acceptability}, the ethical criteria for determining applicable ethical heuristics may itself be problematic.
 \end{enumerate}
In the following sections, we consider how the Four C's analysis of algorithms assists researchers to find points of intersection and incorporation between technical features of algorithms and the extensive literature of ethics, philosophy and moral reasoning.

\section{Part III: Cross-disciplinary results and applications}
In this section, we provide specific examples of how the Four C's Framework can assist in cross-disciplinary understanding of ethical algorithms. We dive deeper into a number of the taxonomic concepts detailed in Part II, in particular providing worked examples and elucidations of how concepts in one field, such as control theory, are usefully deployed in analysis of ethical consequences and regulation.
\subsection{Computability: applications and examples}
As discussed above, the ethical governance of algorithms, from determining ethical classification criteria to devising governance and control frameworks must consider the ways in which technical computability constraints shape their regulation. These considerations impact the extent to which a technical assessment of an algorithm, for example, can be situated or contextualised within particular normative or ethical theories. Part of the challenge is, of course, one of translation and precision. Algorithms, both in their theoretical constitution within computer science and their practical implementation on devices, by and large follow formal computational processes. Logical circuits, subroutines and code are constructed within formal languages, ultimately connecting to (in the classical context and for the most part) discretised assembly and machine code. Putting aside complications arising from undecidability or halting-problem style examples, algorithms follow effective procedures which, while potentially so complex, in theory emerge from the semantics and rules of logic and grammar of the formal systems. Philosophical and ethical criteria which might be used to applied to the study and governance of algorithm, however, exist on a continuum from the fuzzier, vaguer and less formally discernible forms of reasoning within natural language, to attempts to formalise ethics within moral reasoning (see \cite{copp_oxford_2007, richardson_moral_2018}). \\
Incongruence between computational science and ethical criteria has been examined in various contexts, such as terminological vagueness in philosophy and ethics when applied to machine learning \cite{lipton_mythos_2016, lipton_does_2017}. In addition to vagueness, there are also other results from machine learning that bear upon the computational implementation of normative ethics. Among these are so-called \textit{no free lunch} theorems \cite{wolpert_no_1997} which, simplified, provides that any two classifiers are equivalent when performance is averaged across all problems (i.e. in our case, there is no unique universal oracular ethical classifier). Intuitively, this means that classificatory ethical algorithms perform better when different classifiers are applied to different problems. Some ethical classifiers will achieve superior results in some contexts and perform poorly in others, with the consequence of limiting the ubiquity of ethical criteria applied in governance and limiting transferability of ethical insights.\\
\\
Constraints also work in the other direction, that is, from formal results in ethics and moral philosophy to algorithmic design. Various schools of moral and ethical thought, including welfare economics, jurisprudence and philosophy, reveal various \textit{impossibility theorems} which constrain the types of ethical outcomes that are concurrently possible. These results have bearing upon algorithm design and algorithmic governance and should be taken into consideration by technical practitioners in computer science. We provide a few examples.
\begin{enumerate}
    \item \textit{Arrow's impossibility theorem.} One example of a famous impossibility theorems in social choice theory with applicability in the algorithmic domain is in Arrow's impossibility theorem \cite{arrow_difficulty_1950}. The original theorem of Arrow shows that a clear ordering of preferences (according to which, for example utility some other ethical ranking would be based) cannot be determined while adhering to fair voting procedures. Examples of the application of such theorems in algorithmic contexts includes impossibility theorems for clustering \cite{kleinberg_impossibility_2002}, designing group decision procedures \cite{mccomb_impossible_2017}, in swarm algorithms \cite{mogos_voting_2015} and science and engineering more broadly \cite{gaertner_kenneth_2019}.
    \item \textit{Welfare axiology.} A second example of ethical impossibility theorems with applicability to algorithmic governance lies comes from the field of population ethics and welfare axiology in general. One of the celebrated (and much debated) theorems of population ethics is Parfit's `repugnant conclusion'\cite{parfit_reasons_1986}, which one can intuitively interpret as an inconsistency result (that as a result of following an apparently sound set of assumptions one reaches a morally invalid or `repugnant' conclusion so as to give rise to a moral dilemma whose resolution requires either accepting the unsoundness of certain moral assumptions or the logical form of inference used to deduce such a conclusion). Modern approaches, such as \cite{arrhenius_impossibility_2000, arrhenius_impossibility_2011, arrhenius_population_2016} explore in detail the impossibility of consistent ethical outcomes (e.g. the impossibility of a satisfactory population ethics to avoid the repugnant conclusion). Such results have direct application to not only the ethical classification of individual algorithms, but more broadly to how algorithmic ecosystems and networks may be ethically constructed. While one algorithm may meet a specified ethical criteria, it is far from obvious that such criteria extend to other algorithms \textit{and} it is even less obvious how that ethical criteria fares in the presence of interaction effects with other algorithms, say in a distributed network. For these reasons, impossibility results in ethics can help guide ethical algorithmic approaches at the technical and regulatory levels.   
    \item \textit{Utilitarian incompatibility and theories of justice.} Another significant area of ethical and moral (and economic) thought with direct application to algorithmic ethical governance lies social welfare theorems and theories of justice. The canonical issues in this area are exemplified by the ongoing Rawls' \cite{rawls_theory_1999, rawls_justice_2001} and \cite{harsanyi_nonlinear_1975, harsanyi_can_1976, harsanyi_morality_1977} Harsanyi debates about the appropriate utility (social welfare) functions adopted under certain conditions (in short, the debate being about the rationality of and conditions under which certain principles of justice or axiologies would be likely adopted by populations, such as the maxmin rule in social contract settings, or whether principles of average utility would be selected (see \cite{miller_justice_2017} for an overview). Technical details of the debates, such as assumptions over bounded rationality, default preferences (risk aversion or appetite), equiprobability and indifference, have application to both how algorithms are properly classified (i.e. which theories of justice should apply), but also to how they are designed (such as how and whether assumes the principle of indifference and resulting equiprobability of rational choice when algorithms classify). Broader examples of social choice theoretic-results engineered into algorithmic contexts include Rawlsian algorithms for autonomous vehicles \cite{leben_rawlsian_2017}, applying Rawlsian principles to moral agents in a game-theoretic setting \cite{gaudio_connections_2019} and cluster analysis \cite{pierre_barthelemy_median_1981} (see also \cite{gummadi_economic_2019} for a brief example of introducing economic theories of distributive of justice in machine learning).
\end{enumerate}
Some literature in ethical computational science does address the wealth of ethical reasoning about such problems, though on occasions reference to formal ethics is sometimes cursory.

\subsection{Complexity: applications and examples}
Examples of how complexity considerations apply to ethical algorithmic contexts and how complexity theory can itself inform ethical philosophy itself provide a useful illustration of the practical impact that this cornerstone discipline of computer science can have on ethical algorithmic analysis. Examples of computational complexity considerations in an ethical context include the following.
\begin{enumerate}
    \item \textit{Consequentialist and deontological complexity.} Moral philosophy has a well-developed literature covering how constraints on decisional practices, knowledge, resource constraints and bounds on rationality impact the types of normative theories that may be feasibly implemented (see for example \cite{mcnaughton_agent-relativity_1991, savage_foundations_1972}) with particular relevance to computational considerations as to how resource constraints will affect the functioning of ethical algorithms. For example, in \cite{brundage_limitations_2014}, the computational complexity of evaluating the ethical implications of a set of actions given  $N$ agents, $M$ actions to be undertaken in $L$ time is analysed (assessed as $\mathcal{O}(MN^L)$). The work provides an example, albeit simplified, of how complexity considerations impact ethics.
    \item \textit{Strategic and game theoretic contexts}. While not necessarily couched in the language of ethics, economic science overlaps and in many ways formalises many models of ethical behaviour via economic activity of agents and systems. Among many economic subdisciplines, game theory provides an important and vast reservoir of techniques and results of relevance to ethical algorithmic regulation, especially for multi-agent systems. For example, results regarding computationally hard Nash equilibria calculations \cite{conitzer_complexity_2002} have consequences for multi-agent ethical algorithmic systems, where one seeks to find, for example ethically-maximal equilibrium states among algorithmic agents (say representing individuals or users in a policy simulation of optimal resource distribution), or when one seeks to model equilibria for multi-agent algorithms each programmed with distinct ethical criteria (e.g. seeking to find optimal equilibria for self-driving cars each programmed with different ethical constraints). Strategic considerations are arguably central to the broader regulation of algorithms: for example, given a set objective-seeking algorithms, even those encoded with ethical constraints, strategic behaviour will likely emerge as those algorithms seek to optimise for their encoding in ways that are sub-optimal for others. Examples of strategic games of direct application to ethical contexts (and thus relevant to understanding complexity) include cooperative inverse reinforcement learning \cite{hadfield-menell_cooperative_2016} and multi-principal assistance games \cite{fickinger_multi-principal_2020}.
    \item \textit{Fair machine learning.} Fair machine learning (FML), which seeks to mitigate unethical (unfair) algorithmic outcomes (such as bias, discrimination) via fairness criteria, provides an example of a technical sub-discipline at the intersection of computer science and normative ethics where complexity impacts outcomes. FML is a well-developed field with multiple conceptions of fairness (including statistical parity measures of group / subgroup equivalent outcomes, a range of confusion-based metrics, individual fairness measures where similar individuals ought to be subject to similar decisional outcomes and group fairness (see \cite{chouldechova_frontiers_2018, del_barrio_review_2020, caton_fairness_2020} for recent reviews)). The application of computational complexity to ethical FML is an important research direction enabling designers of algorithms to further understand constraints on fairness criteria. A recent example is \cite{kearns_preventing_2018}, where the problem of fairness gerrymandering (how to guarantee statistical fairness across exponentially many subgroups) was examined. Auditing for equality of false positive rates and statistical parity was proven to be equivalent to the computationally hard problem of weak agnostic learning in the asymptotic case and computationally infeasible as the number of subgroups expanded. 
    While near-term specific fairness imperatives may not require such exponential or scaling of the number of protected groups, the result is important for two reasons in particular: (i) the utility of computational complexity results in bounding fairness criteria by illustrating the potentially unavoidable inconsistency of protecting one subgroup while disadvantaging another (e.g. statistical parity for race and gender but adverse impacts for intersectional subroups) and (ii) because imposing fairness constraints should be done with as broad an awareness as possible as to significantly disadvantaged subgroups, in particular, in order to understand trade-offs (see \cite{kearns_empirical_2019} for further discussion). Related work \cite{kleinberg_inherent_2016} on impossibility results that even probabilistic classification of being fair to different groups can be incompatible or potentially uncomputable provide another example of how complexity considerations can assist FML analysis.
\end{enumerate}
An understanding of complexity thus assists practitioners across disciplines with identifying constraints and limitations of algorithmic fairness criteria and, in doing so, assists reasoning about the types of trade-offs which are computationally possible (or even optimal). Explication of trade-offs is an important feature of all ethical algorithmic programs and itself requires a framework to guide investigation. Firstly, not all trade-offs may be known (e.g. there may be uncertainty about which subgroups are adversely affected by an algorithm or would be should fairness constraints be imposed). An example is opacity and uncertainty about the long-term effects of attempting to engineer fairness in particular ways \cite{liu_delayed_2018,zhang_long_2019}. Secondly, there may be no decision-procedure for deciding between trade-offs and indeed more heuristically, there may not be any available explanations or interpretations about how and why algorithms operate in way that allows one to determine the ethical status of an algorithm or computation (needed, for example, in order to decide between competing ethical criteria). Thirdly, the absence of consistent ethical criteria or clear ethical criteria (at least) means falling back upon social norms and practices for informative assumptions and direction \cite{charisi_towards_2017}.

\subsection{Consistency: applications and examples}
The third `C' in the Four C's approach above, consistency, overlaps in many respects with issues discussed in previous sections, especially impossibility results which can be characterised as inconsistency results. In this section, we explore in a bit more detail what is meant by consistency and how algorithms would be audited for ethical consistency. Conceptually, consistency has a deep lineage within the study of formal logical systems, indeed it is in many ways foundational, if by consistency we mean formal consistency. Here we sketch out a few frames within which consistency overall can be considered and articulated by cross-disciplinary researchers.
\begin{enumerate}
    \item \textit{Formal logical consistency.} Formal logical consistency, in various forms that a formal logical system (say of sentential logic) cannot validly result in both a normative proposition $A$ and its negation $\lnot A$ be true (theorems), is an important feature of not only classical formalism in moral philosophy and ethics, but also in algorithmic programming itself. One of the benefits of combining an understanding of computational criteria is that it elucidates the extent to which algorithmic systems can persist with inconsistencies. Researchers in computational science can benefit from considering formal treatments of ethics e.g. in \cite{steele_decision_2020}, while ethical practitioners interested in applications of ethical theories in algorithmic contexts should consider formalisations of such theories (e.g. via logic or even in code).
    \item \textit{Maximal consistency and completeness.} Of course no discussion of formal consistency of ethical algorithms can occur without considering the question of ethical algorithmic completeness (the extent to which every normative proposition or decision of an algorithm must be provably true or false). A second, and following, question arising from consistency considerations is the extent to which all procedures and outcomes of an ethical computation must be self-consistent. This is the ethical algorithmic analogue of maximal consistent sets in formal whether ethical classification of an algorithm requires that it be \textit{maximally ethically consistent}, that is, whether the maximal set of states of the algorithm (in a finite-state context) need to be consistent and complete, or whether it is sufficient that only some subsets of the procedures or the outcomes of a computation need be consistent. Completeness and consistency questions have a vast lineage within logic, philosophy, mathematics and computer science. In this regard, our focus for cross-disciplinary practitioners is more pragmatic than considering the consequences infamous antinomies, such as explicated in G{\"o}del's incompleteness theorems, halting problems and the like for ethical algorithmic practice (the highly specified circumstances of such formally pathological results is of limited relevance to everyday algorithms) though they do resonate, such as via the legacy of Tarski \cite{tarski_decision_1998} in some sense in demonstrating the absence of an absolute grounding upon which to ground normative theories (including any attempts to absolutely ground normative behaviour of algorithms). A practical example is say a typical black-box algorithm which utilises some element of stochasticity such that each implementation of the algorithm is not necessarily reproducible and one does not have guarantees that a computation in one instance is consistent (would not lead to contradiction with) a computation done in another. 
    \item \textit{Approximate consistency.} A third consideration in assessing the ethical status of algorithms is the extent to which approximate consistency can be tolerated. In many ways approximations to consistency requirements present more feasible criteria in practical ethical applications. Thus the tolerances and thresholds for approximate consistency will themselves be a matter for debate. For example, typical metrics in FML, while expressed in terms of equivalence, in reality may require, rather than equiprobability of say group classification, the probability to be within some threshold. An example of this type of approximate fairness the Lipschitz condition on individual fairness explicated in the seminal paper of \cite{dwork_fairness_2012}, where the Lipschitz condition is defined (as is standard) in such as way as to require not identity between metric distance among distributions, but that the difference in such distance be bounded by some constant $K<1$.
\end{enumerate}

\subsubsection{Degrees of consistency} Following on from approximate consistency questions, discussions of consistency in ethical algorithmic contexts can be usefully supplemented via a cross-disciplinary understanding of how consistency and inconsistency in formal, rule-driven protocols, has been managed in social and economic contexts. One must legitimately ask how important is ethical consistency, especially given the diversity of contexts in which ethical algorithms must apply? On the one hand, pure formal consistency, even if the algorithmic form and context were sufficiently well-defined in order to express within formal logical systems, is not feasible for algorithmic systems. On the other hand, some degree of consistency is required by algorithmic regimes: at the extreme end, an algorithm that randomly classified or output random decisions is unlikely to meet any ethical criteria (except perhaps some extreme form of equiprobabilistic parity). Moreover, consistency is required in part for the credibility of algorithms themselves. 

\subsection{Controllability: applications and examples}
The final and fourth consideration from the Four C's approach, controllability, is perhaps one of the most studied and expansive fields across computational science. Controlling automated and mechanical systems has a long history within disciplines such as engineering, computational science and elsewhere for normative approaches to algorithm design and implementation to consider. In this sense, there is already a rich lineage from which algorithm design can draw upon, say in social and economic contexts involving classification that affects rights or resource allocation. To this extent, an understanding of \textit{control theory} is useful for researchers seeking to implement ethical constraints or motivate ethical processes or outcomes among algorithms. Understanding how to control algorithms, correct them and steer them towards consistency with ethical criteria is therefore important. Control theory itself provides a multitude of models of control which ethical algorithm designers can draw upon. A particularly useful approach is in the concept of Kalman filtering where a system is fully controllable at time $T>0$ if it can be transitioned from any initial state to any other target state in time $T$ by a choice of controls (see \cite{dorf_modern_2011} for a general discussion). In an ethical context, this can be framed in terms of a requirement that those states transitioned between meet some ethical criteria (e.g. the system does not enter a state in which it inappropriately uses or represents a protected attribute) and that the set of controls themselves meets any relevant ethical criteria. Furthermore, one can frame the question of where and how to intervene in order to adequately control algorithms via categorisations from within FML categorisation within FML literature \cite{caton_fairness_2020}. Control may be imposed at the following stages: (i) preprocessing, (ii) in-processing (within the model); and (iii) post-processing. We briefly sketch examples of each type of control in an ethical algorithmic context.

\begin{enumerate}
    \item \textit{Preprocessing.} Here the idea is to impose ethical criteria on the inputs to an algorithm in order to increase the probability that the algorithm satisfies ethical criteria. A typical example includes methods to \textit{redact} certain protected attributes (e.g. race or class) \cite{mcnamara_provably_2017}. However, redaction can lead to declines in accuracy, in a formal control sense, the application of controls during a preprocessing state diminishes the transition probability from initial states to optimal states. The point here is, however, that framing algorithms in terms of controls can usefully assist program design to meet ethical criteria. One of the challenges of preprocessing is the ability of data to relearn protected attributes, such as in deep neural networks where intermediate feature layers may effectively reconstruct redacted attributes due to high correlation of those attributes with other features (demonstrating the importance of understanding the algorithmic engineering and effects when seeking to impose such deontological criteria). 
    \item \textit{In-processing.} Controlling algorithms during model execution of `in-flight', either using open-loop or closed-loop control, is an important consideration in ethical algorithm design. Most algorithms are executed on static code or at least code that does not dynamically vary with each execution.  Controls tend to occur via feedback in ways that are ultimately connected to achieving some objective, such as optimising a task and minimising loss of some description. Thus in FML contexts, typical in-processing control occurs by imposing conditions upon the relevant loss function (such as requiring certain equiprobabilities of outcomes or ratios etc drawn from confusion matrices \cite{caton_fairness_2020}), which, in a deep learning context, usually then relies upon the implicit control via backpropagation that steers algorithms in desired directions via updating weights. 
    \item \textit{Post-processing.} Post-processing concerns adjusting the outcomes of algorithms once run in order to meet certain ethical or other criteria. In some senses the ability to exert such a degree of control implies a ubiquitous level of control on the ultimate outcome of algorithmic processes itself. Ethically, one consideration for researchers is the morality of adjusting outcomes should, for example, the algorithm have satisfied ethical deontological criteria (i.e. if the process was considered ethical, is post-processing itself ethical)? 
\end{enumerate}
Of course, as with any engineering design, designing algorithms for ethical control faces a range of potential challenges: (a) \textit{non-linearity}, where inputs may combine in a non-linear way or the functional form of an algorithm may depend upon inputs in an unpredictable way or conditional upon data (e.g. removing features may vary functional form). In this case, direct attempts to ethically control algorithms via hard-coding requirements may be frustrated by inherent uncertainty or even stochasticity in subroutine execution; (b) \textit{non-repeatability}, again where in particular stochastic processes are involved in algorithms (say random walk-based algorithms), it may be difficult to adequately control the algorithm's direction for any one iteration; (c) \textit{system complexity}, it may be difficult to map a limited set of simple controls to complex systems and indeed controllability itself may require a separate class of algorithms whose exact control regime is to some degree unknown; (d) \textit{proximity}, controllability may not equate to full control, that is, a system may be ethically steered to within some tolerance or threshold $\epsilon$ of the desired ethical state. Understanding acceptable proximity versus idealised outcomes from an ethical perspective is crucial, especially in any practical attempt to ethically confirm algorithms.

\section{Conclusions: responses and ways forward}
Of course, having identified key challenges in ethical algorithmic research, such as appropriate grounds upon which to decide between competing ethical criteria, how to manage and tolerate uncertainty and inconsistency and vagueness, it begs the question to ask how such potential conundrums may be addressed. One avenue, and in some ways a necessary one given that ultimately ethical regulation of algorithms is via law, is to analyse how legislative and administrative legal regimes handle such ambiguity. For example, ethical governance may adopt the type of epistemological framing the law adopts in distinguishing \textit{legal} from \textit{equitable} jurisprudence and remedies. For example, by analaogy, the legal aspect of ethical algorithmic engineering (as distinct from the legislative regimes that may apply to algorithms) could involve separating coding constraints into (i) \textit{hard-coded} constraints e.g. hard-coding canonical protected attributes, but allowing (ii) an equity-style deviation from the legal norm where the facts or context differ. Another option mooted in the literature to deal with moral dilemmas is to engineer ambiguity, or `uncertainty by design' \cite{bogosian_implementation_2017}, where instead of a shortcoming, uncertainty and ambiguity are themselves effectively treated as \textit{resources} that allow algorithms to adapt more effectively to changed circumstances, albeit with higher risks of inconsistency. On the FML front, a range of proposals (see \cite{del_barrio_review_2020} for example) exist to either minimise or reconcile inconsistencies between individual and statistical fairness (see \cite{kearns_preventing_2018}) on rich subgroup fairness where the authors assert such reconciliation is possible when the VC dimension of the classification model is sufficiently bounded).

We hope that the Four C's Framework and characterisations of ethical algorithms which we have presented in this paper for use by practitioners across research fields in order to assist them with cross-disciplinary analysis of the ethics of algorithmic systems will provide intelligible examples of how reaching across the disciplines can assist in motivating solutions to ethical problems. In this spirit, to address these computational challenges, one may reach out to other disciplines to see how similar phenomena have been dealt with. By firstly clearly situating the ethics of algorithms within formal computational understanding of what algorithms are, together with key features of computability, complexity, consistency (and completeness) and controllability, researchers from a variety of disciplines may then work within a coherent framework enabling comparison and translation of of analytical methods and results in a way that may lead to the types of synergistic solutions for  cross-disciplinary research is uniquely positioned to provide.

\newpage

\bigskip

\bibliography{references.bib}
\end{document}